# Photonic ferromagnetic-like spontaneous mode-locking phase transition with replica symmetry breaking in multimode Nd:YAG laser

André L. Moura[1,2], Pablo I. R. Pincheira[1], Ernesto P. Raposo[3], Anderson S. L. Gomes[1] & Cid B. de Araújo[1]

**The recent reports of the replica symmetry breaking (RSB) phenomenon in photonic experiments [1-5] boosted the understanding of the role of disorder in multimode lasers, as well as helped to settle enlightening connections [6-13] with the statistical physics of complex systems. RSB manifests when identically-prepared system replicas reach distinct states, yielding different measures of observable quantities [14]. Here we demonstrate the RSB in the spontaneous mode-locking regime of a *conventional* multimode Nd:YAG laser in a closed cavity. The underlying mechanism is quite distinct from that of the RSB spin-glass phase in cavityless random lasers with incoherently-oscillating modes. Here, a specific nonuniform distribution of the gain takes place in each pulse, and frustration is induced since the coherent oscillation of a given subset of longitudinal modes dominates and simultaneously inhibits the others. Nevertheless, when high losses are introduced only the replica-symmetric amplified stimulation emission is observed. We therefore suggest that the RSB transition can be used as an identifier of the threshold in standard multimode lasers, as recently proposed and demonstrated for random lasers [1,2].**

    The concept of RSB appeared in the late 1970's in the context of Parisi's theory of disordered magnetic systems [14,15]. In this framework, for sufficiently low temperatures and strong disorder, the free energy landscape breaks into a large number of local minima in the configuration space. Due to frustrated magnetic interactions in the disordered Hamiltonian, spins fail to align in a spatially regular configuration, as in the ferromagnetic state. Instead, spins "freeze" along random directions in a *spin-glass* state. As a given spin configuration can be trapped in a local free energy minimum, metastability and irreversibility effects arise in the spin-glass phase. Consequently, identical systems prepared under identical conditions (system replicas) can reach different states with different measures of observable quantities and nontrivial correlation patterns in a *replica symmetry breaking* scenario. Later on, the scope of this concept was much extended to reach other complex systems [14], including neural networks and structural glasses.

    In the photonic context, theoretical predictions of RSB behavior emerged in the last decade [6-13], mainly related to the properties of multimode random lasers (RLs). In the photonic-to-magnetic analogy, the mode amplitudes play the role of the spins and the excitation pumping energy acts as the inverse temperature. In 2015, the very first experimental evidence of photonic RSB arose in a two-dimensional (2D) functionalized $T_5OC_x$ oligomer amorphous solid-state RL [1]. Subsequent demonstrations of RSB appeared in 1D erbium-doped random fiber laser [4,5] and 3D functionalized $TiO_2$ particle-based dye-colloidal [3] and neodymium-doped $YBO_3$ solid-state [2] RLs. In these cavityless systems, the manifestation of RSB concurs with the settlement of a glassy phase of light – a photonic spin-glass phase, in which the modes are nontrivially correlated though displaying incoherent oscillation [6-13]. The threshold separating the prelasing regime with amplified spontaneous emission (ASE) and the RL behavior thus

[1]Departamento de Física, Universidade Federal de Pernambuco, 50670-901, Recife-PE, Brazil. [2]Grupo de Física da Matéria Condensada, Núcleo de Ciências Exatas - NCEx, Campus Arapiraca, Universidade Federal de Alagoas, 57309-005, Arapiraca-AL, Brazil. [3]Laboratório de Física Teórica e Computacional, Departamento de Física, Universidade Federal de Pernambuco, 50670-901, Recife-PE, Brazil.

became identified with the phase transition from the photonic paramagnetic-like to the glassy regime. Remarkably, the theoretical analysis also predicted [6,12] the possibility of RSB in the mode-locking laser regime with weak disorder. The magnetic analogue of this phase, with most modes oscillating coherently, is the *random bond ferromagnet* with RSB, whose free energy landscape also presents a large number of local minima and configurations displaying domains of disordered spins that nucleate in a ferromagnetic background of aligned spins [16,17].

Disorder and frustration are central ingredients to the occurrence of RSB [14]. The most usual form to introduce disorder and frustration is through quenched random couplings, either between spins or active modes. Nevertheless, on the photonic side a recent instigating work [18] demonstrated the RSB in the threshold region of *conventional* solid-state and liquid multimode lasers, associated with the onset of strong intensity fluctuations and without any form of intentionally added disorder. These systems contrast markedly with the RSB glassy RLs, in which the disorder strength depends on the intrinsic properties of the nonlinear random medium and photonic frustration is induced due to the impossibility of satisfying every random coupling [6-13].

In the present work, we report on the evidence of RSB in the spontaneous mode-locking (SML) laser regime of a *standard* Q-switched Nd:YAG pulsed laser in a closed cavity, pumped by a flash lamp (see Fig. 1 and Methods). For each value of the voltage $V_{app}$ applied to the flash lamp, data from 500 spectra were recorded. The multimode character of the laser system arises from the large cavity of length ~1 m. We recall that the SML laser regime was observed in a RL consisting of micrometer-sized cluster of titania nanoparticles suspended in a rhodamine dye solution, but the occurrence of RSB was not reported [19], despite theoretical predictions (see [6] and references therein).

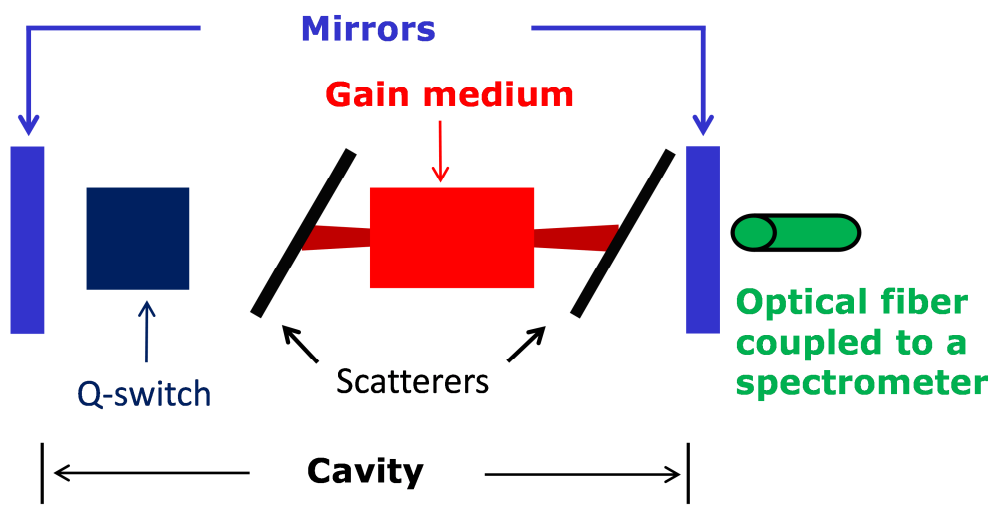

**Figure 1 | Experimental setup.** Q-switched Nd:YAG pulsed laser with single-shot spectra collected by a multimode optical fiber coupled to a spectrometer. A large cavity of length ~1 m gives rise to the multimode character of the laser system. If high losses are introduced by the presence of scatterers between the crystal and the cavity mirrors, laser oscillation is suppressed.

In Fig. 2 the mean total emitted intensity and linewidth reduction, inferred by the full width at half maximum (FWHM), are plotted as a function of $V_{app}$. The abrupt change in FWHM (blue circles) fixes the threshold value at 1.01 kV. On the other hand, as $V_{app}$ increases, the bandwidth narrowing and higher intensity peaks (see spectra in Fig. 3) makes the emitted intensity (red triangles) to enhance smoothly with the applied voltage.

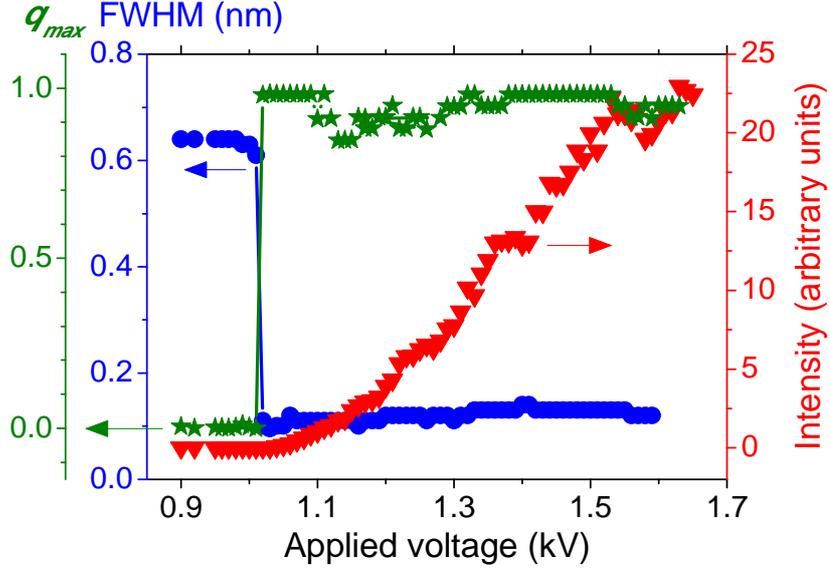

**Figure 2 | RSB transition from the replica-symmetric ASE to the RSB SML phase in the conventional multimode Nd:YAG laser.** Total emitted intensity (red triangles), averaged over 500 spectra for each applied voltage, FWHM (blue circles), and parameter $q_{max}$ (green stars) at which the distribution P($q$) of replica overlaps of shot-to-shot intensity fluctuations is maximum as a function of $V_{app}$. The abrupt change in FWHM at the threshold from the ASE to the SML laser regime remarkably coincides with the sharp transition from the replica-symmetric ($q_{max} \sim 0$) to the RSB ($q_{max} \sim 1$) regime.

The identification of RSB in a photonic system relies [6-13] on the analysis of the mode correlations among distinct replicas, which can be calculated by the replica overlap parameter [1,6]

$$q_{\gamma\beta} = \frac{\sum_k \Delta_\gamma(k) \Delta_\beta(k)}{\sqrt{\left[\sum_k \Delta_\gamma^2(k)\right]\left[\sum_k \Delta_\beta^2(k)\right]}}, \qquad (1)$$

where $\gamma, \beta = 1, 2, \ldots, N_s$ denote the replica labels ($N_s = 500$ in the present work), the mean intensity at the wavelength indexed by $k$ reads $\langle I \rangle(k) = \sum_{\gamma=1}^{N_s} I_\gamma(k)/N_s$, and the intensity fluctuation is $\Delta_\gamma(k) = I_\gamma(k) - \langle I \rangle(k)$. In the photonic context, each intensity spectrum generated by a pulse of the flash lamp is considered a *replica*, i.e. a copy of the system under fairly identical experimental conditions. The probability density function (PDF) P($q$), analogue to the Parisi order parameter in spin-glass theory [14], describes the distribution of replica overlaps $q = q_{\gamma\beta}$, signaling a replica-symmetric or a RSB phase if it peaks exclusively at $q_{max} = 0$ or also at values $q_{max} \neq 0$, respectively.

We show the PDFs P($q$) in Fig. 3, along with the respective profiles of 20 emitted spectra at each $V_{app}$. The indicated voltages are representative of the prelasing (0.90 kV), threshold (1.01 kV), nearly above threshold (1.02 kV), and SML laser (1.63 kV) regimes. In the former two cases, the replica-symmetric distributions displaying a single peak at $q_{max} = 0$ correspond to the ASE behavior with weak intensity fluctuations. Right above the threshold, the PDF P($q$) starts to present a double-peaked pattern in the SML laser regime, with $q_{max} \neq 0$, thus signaling the onset of the RSB phase with strong fluctuations and full suppression of the fluorescence maximum. In Fig. 2 we also notice the remarkable coincidence between the abrupt change in FWHM (blue circles) at the threshold from the ASE to the SML laser behavior and the sharp transition from the replica-symmetric ($q_{max} \sim 0$) to the RSB ($q_{max} \sim 1$) regime (green stars).

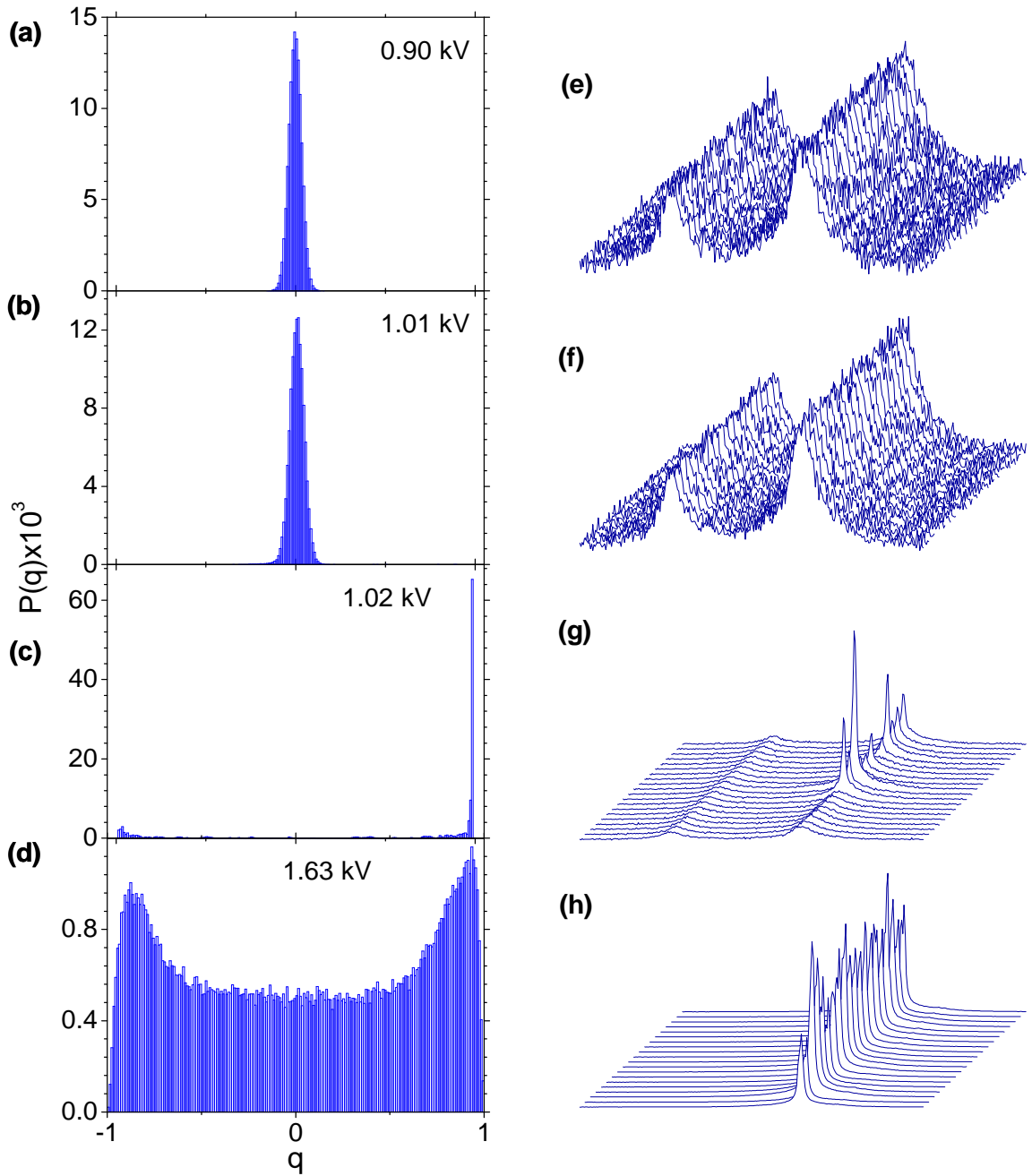

**Figure 3 | Pulse-to-pulse intensity fluctuations and corresponding overlap distributions signaling the RSB transition in the multimode Nd:YAG laser system.** (a)-(d) PDFs P($q$) of replica overlaps of shot-to-shot intensity fluctuations for applied voltages in the prelasing ASE (0.90 kV), threshold (1.01 kV), nearly above threshold (1.02 kV), and SML laser (1.63 kV) regimes. Replica-symmetric and RSB behaviors are respectively related to distributions displaying a single peak at $q_{max} = 0$ and double peaks with $q_{max} \neq 0$. (e)-(h) 20 emission spectra at each $V_{app}$ associated with the PDFs shown in (a)-(d). Intensity fluctuations grow with the applied voltage and become rather strong in the RSB SML laser regime.

The above conclusions are strengthened by the experiment in which high losses are introduced by placing two pieces of paper playing the role of efficient scatterers between the crystal and the cavity mirrors of the very same Nd:YAG laser device, shown in Fig. 1. As laser oscillations are suppressed in this case, the SML laser regime cannot establish and the ASE behavior takes place for all applied voltages, with absence of threshold (see Fig. 4). Consistently, a replica-symmetric picture clearly emerges for all $V_{app}$ values, with distributions P($q$) displaying only a single peak at $q_{max} = 0$, as shown in Fig. 5. Interestingly, the observed broadening of the PDFs P($q$) is associated with an increase in the intensity fluctuations, which remain, however, still much weaker than in the lasing regime. The lower fluorescence maximum sustains even for high voltages.

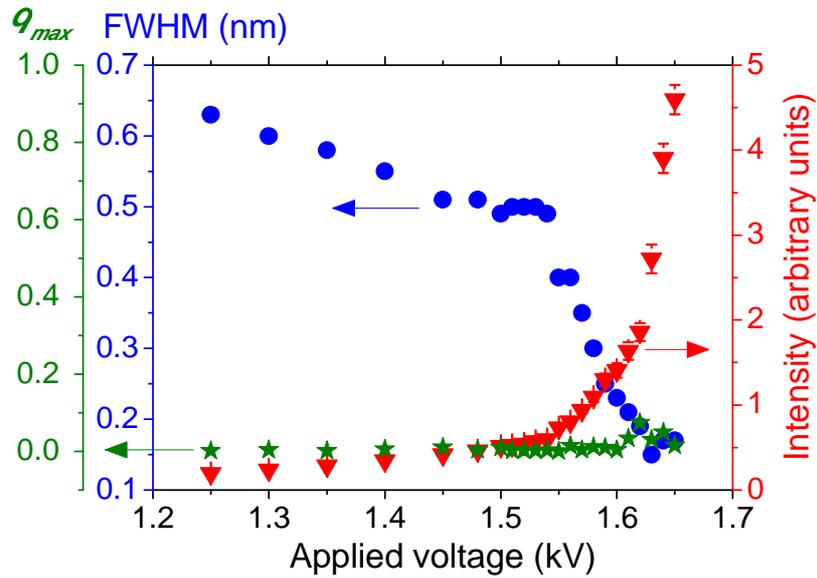

**Figure 4 | Absence of lasing threshold and replica-symmetric ASE regime with high losses.** Quantities as in Fig. 2 when high losses are introduced to prevent laser oscillation. No threshold to the SML laser regime is observed in the behavior of FWHM, as the replica-symmetric ($q_{max} \sim 0$) ASE regime emerges for all applied voltages.

We therefore argue that the transition from the replica-symmetric to the RSB phase could be used as an identifier of the threshold in standard multimode lasers. This suggestion, which was recently proposed and demonstrated in the context of random lasers [1,2], also holds some analogy with the proposal [20] of applying the Lévy exponent α, which describes the statistics of the emitted intensities [20-22], to determine the transition from the Gaussian prelasing to the Lévy RL behavior in RL systems.

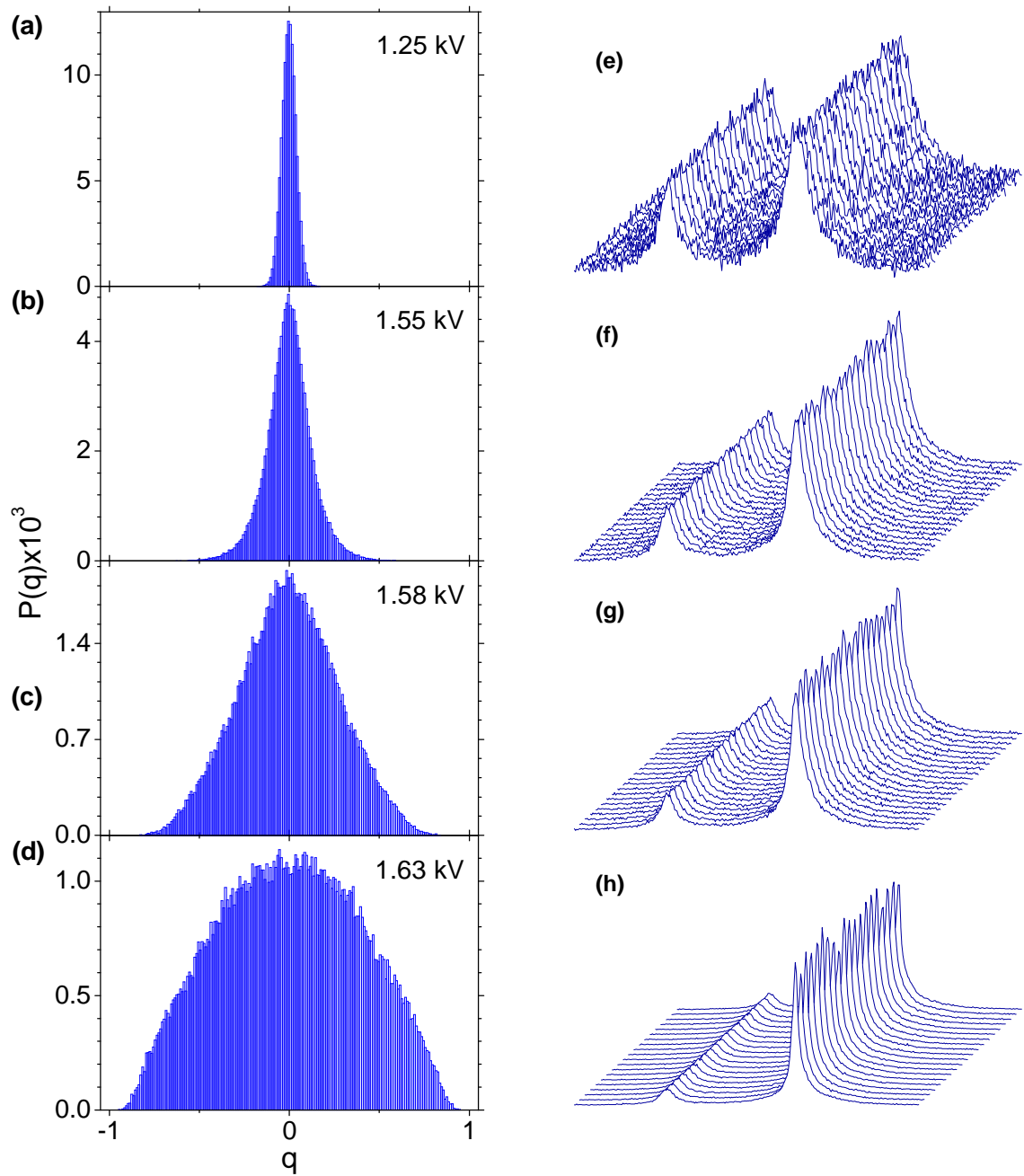

**Figure 5 | Pulse-to-pulse intensity fluctuations and corresponding overlap distributions evidencing the absence of transition and the replica-symmetric ASE regime.** Description as in Fig. 4 when high losses are introduced to prevent laser oscillation. (a)-(d) PDFs P($q$) show a single peak at $q_{max} = 0$ for all applied voltages, as characteristic of the replica-symmetric ASE regime. (e)-(h) Intensity fluctuations grow with the applied voltage but still remain much weaker than those in the RSB SML laser regime at high voltages. The lower fluorescence maximum is not suppressed in this ASE regime.

At this point, one relevant issue is to understand the origin of the disorder and frustration mechanisms underlying the RSB picture in the SML laser regime of the standard multimode Nd:YAG laser device. As mentioned, RSB cannot sustain in a system with complete absence of these ingredients. At each excitation pulse, a specific set of spatially superposed longitudinal modes establishes. As they intrinsically compete for the gain, photonic frustration is induced in the sense that, in each shot, the coherent oscillation of a given subset of coupled modes dominates and simultaneously inhibits the coherent oscillation of others. As a consequence, strong shot-to-shot intensity fluctuations emerge since each shot yields a specific nonuniform distribution of the gain among the active modes. In the statistical physics language, each excitation pulse emitted under the same experimental conditions leads the associated photonic replica to occupy a distinct thermodynamic state with a particular pattern of intensity spectrum. The characterization of the resulting shot-to-shot intensity fluctuations through the replica overlap parameter, equation (1), over an extensive sequence of pulses (large number of replicas) therefore demonstrates the emergence of the RSB scenario at and above the threshold to the multimode SML laser regime.

We also comment on the important contrast between this finding and the recent observation of the transition from the replica-symmetric prelasing to the RSB spin-glass phase in disordered RL systems [1-5]. Indeed, though RSB is identified above the threshold in both multimode SML laser and RL regimes, we recall that the photonic phases are actually quite distinct, with most modes oscillating coherently in the former, whereas in the latter they present nontrivial correlations and incoherent oscillation [6-13]. In this sense, our work can also possibly represent the first experimental demonstration of the photonic RSB SML laser regime theoretically predicted in [6,12].

In conclusion, in this work a conventional multimode Nd:YAG laser in a closed cavity was employed to unveil the RSB phase at and above the threshold to the SML laser regime. The photonic mechanism underlying this behavior is quite distinct from that related to the RSB glassy phase in RLs. The signature of RSB has been clearly identified in the distribution of replica overlaps of pulse-to-pulse intensity fluctuations. In addition, the introduction of high losses in the cavity has led to the observation of only the ASE regime with replica symmetry. We thus suggest that the transition from the replica-symmetric to the RSB behavior could be used as an identifier of the lasing threshold in standard multimode lasers.

**Methods**

Experiments were performed using a commercial Q-switched Nd:YAG pulsed laser, with emission at 1064 nm due to the $Nd^{3+}$ $^4F_{3/2} \rightarrow {}^4I_{11/2}$ transition, 5 Hz repetition rate, and 7 ns pulse duration. The experimental setup showing the laser cavity of length ~1 m is depicted in Fig. 1. Single-shot spectra were collected by a multimode optical fiber coupled to a spectrometer with nominal resolution of 0.2 nm. For each value of the voltage $V_{app}$ applied to the flash lamp to excite the Nd:YAG crystal, data from 500 spectra were recorded and analyzed.

**Acknowledgements**

We thank the PRONEX Program, National Institute of Photonics – INFO, CNPq, FACEPE, CAPES, and Finep (Brazilian agencies) for the financial support.

**Author Contributions**

A.L.M., A.S.L.G., E.P.R. and C.B.A. conceived and designed the study. A.L.M. and P.I.R.P. performed the experiments. A.L.M., P.I.R.P., A.S.L.G., E.P.R. and C.B.A. discussed the results and wrote the manuscript.